\documentclass[dvips]{acta}
\usepackage{supertabular,lscape,epsfig}
\usepackage{amssymb}
\usepackage{amsmath}
\DeclareSymbolFont{ppa}{OT1}{ppl}{m}{it}
\DeclareMathSymbol{\vv}{\mathalpha}{ppa}{'166}

\SetPages{163}{185}

\SetVol{58}{2008}

\begin{document}

\newcommand{\TabCapp}[2]{\begin{center}\parbox[t]{#1}{\centerline{
  \small {\spaceskip 2pt plus 1pt minus 1pt T a b l e}
  \refstepcounter{table}\thetable}
  \vskip2mm
  \centerline{\footnotesize #2}}
  \vskip3mm
\end{center}}

\newcommand{\TTabCap}[3]{\begin{center}\parbox[t]{#1}{\centerline{
  \small {\spaceskip 2pt plus 1pt minus 1pt T a b l e}
  \refstepcounter{table}\thetable}
  \vskip2mm
  \centerline{\footnotesize #2}
  \centerline{\footnotesize #3}}
  \vskip1mm
\end{center}}

\newcommand{\MakeTableSepp}[4]{\begin{table}[p]\TabCapp{#2}{#3}
  \begin{center} \TableFont \begin{tabular}{#1} #4
  \end{tabular}\end{center}\end{table}}

\newcommand{\MakeTableee}[4]{\begin{table}[htb]\TabCapp{#2}{#3}
  \begin{center} \TableFont \begin{tabular}{#1} #4
  \end{tabular}\end{center}\end{table}}

\newcommand{\MakeTablee}[5]{\begin{table}[htb]\TTabCap{#2}{#3}{#4}
  \begin{center} \TableFont \begin{tabular}{#1} #5
  \end{tabular}\end{center}\end{table}}

\newfont{\bb}{ptmbi8t at 12pt}
\newfont{\bbb}{cmbxti10}
\newfont{\bbbb}{cmbxti10 at 9pt}
\newcommand{\uprule}{\rule{0pt}{2.5ex}}
\newcommand{\douprule}{\rule[-2ex]{0pt}{4.5ex}}
\newcommand{\dorule}{\rule[-2ex]{0pt}{2ex}}
\def\thefootnote{\fnsymbol{footnote}}

\hyphenation{Ce-phe-ids eclip-ses chan-ges}

\begin{Titlepage}
\Title{The Optical Gravitational Lensing Experiment.\\
The OGLE-III Catalog of Variable Stars.\\
I. Classical Cepheids in the Large Magellanic Cloud\footnote{Based on
observations obtained with the 1.3-m Warsaw telescope at the Las Campanas
Observatory of the Carnegie Institution of Washington.}}
\Author{I.~~S~o~s~z~y~\'n~s~k~i$^1$,~~
R.~~P~o~l~e~s~k~i$^1$,~~
A.~~U~d~a~l~s~k~i$^1$,~~
M.\,K.~~S~z~y~m~a~{\'n}~s~k~i$^1$,\\
M.~~K~u~b~i~a~k$^1$,~~
G.~~P~i~e~t~r~z~y~\'n~s~k~i$^{1,2}$,~~
\L.~~W~y~r~z~y~k~o~w~s~k~i$^{1,3}$,\\
O.~~S~z~e~w~c~z~y~k$^{1,2}$~~
and~~K.~~U~l~a~c~z~y~k$^1$}
{$^1$Warsaw University Observatory, Al.~Ujazdowskie~4, 00-478~Warszawa, Poland\\
e-mail:
(soszynsk,rpoleski,udalski,msz,mk,pietrzyn,wyrzykow,szewczyk,kulaczyk)
@astrouw.edu.pl\\
$^2$ Universidad de Concepci{\'o}n, Departamento de Fisica, Casilla 160--C,
Concepci{\'o}n, Chile\\  
$^3$ Institute of Astronomy, University of
Cambridge, Madingley Road, Cambridge CB3 0HA, UK}
\Received{August 15, 2008}
\end{Titlepage}
\Abstract{We present the first part of a new catalog of variable stars 
(OIII-CVS) compiled from the data collected in the course of the third
phase of the Optical Gravitational Lensing Experiment (OGLE-III). In this
paper we describe the catalog of 3361 classical Cepheids detected in the
$\approx40$ square degrees area in the Large Magellanic Cloud. The sample
consists of 1848 fundamental-mode (F), 1228 first-overtone (1O), 14
second-overtone (2O), 61 double-mode F/1O, 203 double-mode 1O/2O, 2
double-mode 1O/3O, and 5 triple-mode classical Cepheids. This sample is
supplemented by the list of 23 ultra-low amplitude variable stars which may
be Cepheids entering or exiting instability strip.

The catalog data include {\it VI} high-quality photometry collected since
2001, and for some stars supplemented by the OGLE-II photometry obtained
between 1997 and 2000. We provide basic parameters of the stars:
coordinates, periods, mean magnitudes, amplitudes and parameters of the
Fourier light curve decompositions. Our sample of Cepheids is
cross-identified with previously published catalogs of these variables in
the LMC. Individual objects of particular interest are discussed, including
single-mode second-overtone Cepheids, multiperiodic pulsators with unusual
period ratios or Cepheids in eclipsing binary systems.

We discuss the variations of the Fourier coefficients with periods and
point out on the sharp feature for periods around 0.35~days of
first-overtone Cepheids, which can be explained by the occurrence of 2:1
resonance between the first and fifth overtones. Similar behavior at
$P\approx3$~days for 1O Cepheids and $P\approx10$~days for F Cepheids are
also interpreted as an effect of resonances between two radial modes. We
fit the period--luminosity relations to our sample of Cepheids and compare
these functions with previous determinations.}{Cepheids -- Stars:
oscillations -- Magellanic Clouds}

\Section{Introduction}
The Optical Gravitational Lensing Experiment (OGLE) is a wide-field sky
survey originally motivated by search for microlensing events
(Paczy{\'n}ski 1986). The observing strategy of the project is to regularly
monitor brightness of about 200~million stars in the Magellanic Clouds and
Galactic bulge in the time-scales of years. A~by-product of these
observations is an enormous database of photometric measurements, which can
be used for selecting long lists of newly discovered variable stars.

The OGLE project yielded a wealth of information about variable stars. The
second phase of the survey, conducted between 1997 and 2000, resulted in
catalogs of thousands Cepheids, RR~Lyr stars, eclipsing binaries and long
period variables in the Magellanic Clouds. Moreover, the huge catalogs of
variable sources found in the OGLE-II fields in the Magellanic Clouds
({\.Z}ebru{\'n} \etal 2001) and in the Galactic bulge (Wo{\'z}niak \etal
2002) were released. In this paper we present the first part of the
OGLE-III Catalog of Variable Stars (OIII-CVS) -- the catalog containing
virtually all variable stars in the fields regularly observed by OGLE since
2001 with the Warsaw telescope at Las Campanas Observatory, Chile.

Classical Cepheids ($\delta$~Cep stars, type I Cepheids), as the primary
distance indicator, are among the most important variable stars. The Large
Magellanic Cloud (LMC) is one of the most fundamental extragalactic targets
of modern astrophysics, because it is our nearest non-dwarf neighbor
galaxy. For this reason we begin the OIII-CVS with the catalog of classical
Cepheids in the LMC.

Large number of variable stars in the LMC, including classical Cepheids,
were discovered by Leavitt (1908). However, the first period derivation and
the plot of the period--luminosity (PL) diagram for 40 LMC Cepheids was
made by Shapley (1931). In 1955 the periods of 550 classical Cepheids in
the LMC were published (see Shapley and McKibben Nail 1955 for the
bibliography). Then, a considerable survey for LMC Cepheids was done by
Woolley \etal (1962). The catalog prepared by Payne-Gaposchkin (1971) on
the basis of Harvard photographic plates contained about 1100 Cepheids in
the LMC. After hiatus, a number of Cepheids in the LMC were also discovered
by Hodge and Lee (1984), Kurochkin \etal (1989), van Genderen and Hadiyanto
Nitihardjo (1989) and Mateo \etal (1990). In the late 1990's very large
catalogs of Cepheids were published as a by-product of gravitational
microlensing surveys: EROS (Beaulieu \etal 1995, Afonso \etal 1999), MACHO
(Welch \etal 1997, Alcock \etal 1999b) and OGLE-II (Udalski \etal 1999d,
Soszy{\'n}ski \etal 2000).

The catalog described in this work contains the largest sample of classical
Cepheids detected to date in the LMC and, likely, in any other
environment. Almost 1000 objects are new identifications. Double-mode
Cepheid sample presented in this paper is three times more numerous than
the largest sample presented so far. We also show individual objects of
particular interest, like triple-mode Cepheids, Cepheids with non-radial
pulsations, Cepheids in eclipsing binary systems, etc.

The paper is organized as follows. In Section~2 we describe how the
observations were obtained and reduced. Section~3 gives the details about
the process of Cepheid selection. In Section~4 we describe the catalog
itself. We compare our sample with previously published catalogs of the LMC
classical Cepheids in Section~5. In Section~6 we discuss the Fourier
coefficients as a function of periods. In Section~7 we fit the
period--luminosity relations. Finally, Section~8 summarizes the paper.

\Section{Observations and Data Reduction}
The photometric data were obtained with the 1.3-m Warsaw telescope located
at Las Campanas Observatory in Chile. The observatory is operated by the
Carnegie Institution of Washington. The telescope is equipped with the
``second generation'' camera consisting of eight SITe $2048\times4096$ CCD
detectors with 15~$\mu$m pixels what corresponds to 0.26 arcsec/pixel
scale. The gain of the chips is adjusted to be about 1.3
$\mathrm{e}^-/\mathrm{ADU}$ with the readout noise from 6 to 9
$\mathrm{e}^-$ depending on the chip. For details of the instrumental setup
we refer to Udalski (2003).

Observations of 116 OGLE-III fields covering 39.7 square degrees of the LMC
started in July 2001. The data presented in this paper were collected up to
March 2008. In the future, photometry provided with the catalog will be
supplemented by observations obtained after this date, up to the end of the
third phase of the OGLE survey.

The photometry was obtained using Difference Image Analysis (DIA) technique
(Alard and Lupton 1998, Alard 2000, Wo{\'z}niak 2000), which is able to
perform in dense stellar fields considerably better photometry than the
traditional PSF-fitting programs. We emphasize that even though there are
small gaps between the chips of the CCD mosaic, our final DIA photometry
pipeline (Udalski \etal 2008a) provides photometry of stars from the entire
fields, because, due to imperfections of the telescope pointing, the
regions between the chips are also observed from time to time. Thus, the
completeness of the catalog is practically not limited by the lack of
observations in the gaps between the chips, although the smaller number of
points at the edges of the fields sometimes may decrease the efficiency of
variability search.

About 90\% of observations were taken in the $I$ photometric band closely
resembling standard filter. The remaining frames were observed in the $V$
bandpass. The data were calibrated to the standard system using hundreds of
thousands stars observed during the OGLE-II project (1997--2000) and common
with the OGLE-III observations. The procedure of calibration of the data
and astrometric transformations were described by Udalski \etal
(2008a). The accuracy of the photometric calibrations is better than
0.02~mag, while equatorial coordinates in the catalog have uncertainties of
about 0\zdot\arcs1.

Photometry error bars derived by the DIA package were known to be
underestimated. Here we corrected the error bars using technique derived
for OGLE-II LMC microlensing events search and described in detail in
Wyrzykowski \etal (in preparation). In brief, the method compares observed
photometric scatter of constant stars in a given field with their mean
error bars and fits the coefficients of relation between original and
corrected error bar: $\sigma_{\rm corr} = ((\gamma\sigma)^2+
\epsilon^2)^{1/2}$. For OGLE-III data $\gamma$ and $\epsilon$ were derived
independently for each field and CCD chip. On the average $\gamma=1.204$
and $\epsilon=0.0046$ in the $I$ band. In $V$ filter: $\gamma=0.996$ and
$\epsilon=0.0035$.

For the central 4.5 square degrees of the LMC the OGLE-II photometry
(Szyma{\'n}ski 2005) was available. When it was possible we tied both
datasets to obtain the time base of observations covering 12 years. For
each star we shifted the OGLE-II photometry to agree with the mean OGLE-III
magnitudes, however in some peculiar Cepheids with variable mean magnitudes
more advanced procedure should be applied.

Our sample contains 47 Cepheids with no {\it I}-band data in the OGLE-III
database, usually due to exceeding the CCD saturation limit (which is about
$I=13$~mag). In a few cases our photometric techniques failed in the
centers of the clusters. For 16 of these objects the OGLE-II {\it I}-band
photometry is available, where the level of saturation was slightly higher
($I=12.5$~mag). Note, that the OGLE-III photometry for the brightest stars
is sometimes more noisy due to exceeding the saturation limit.

Luckily, most of these bright or cluster stars have observations in the $V$
filter, because our {\it V}-band data for classical Cepheids saturate for
periods longer than $\approx50$~days. Only for the five longest-period
Cepheids, neither $I$ nor {\it V}-band photometry are available. However,
even for these stars we could measure periods and shapes of their light
curves, because in the DIA technique every bright variable star produces a
number of artificial objects in the closest neighborhood that mimic the
variability of this bright star. This is because the DIA method does not
subtract the profiles of the neighboring objects while doing photometry. We
used these artificial light curves for measuring periods of the long-period
Cepheids. The calibrated {\it VI} photometry for brightest stars in the LMC
will be published soon, when the shallow survey conducted on the Warsaw
telescope will be finished. At that moment we will be able to supplement
our catalog with the data for the brightest Cepheids.

\Section{Selection of Cepheids}
\subsection{Single-Mode Cepheids}
The search for classical Cepheids in the LMC was preceded by a massive
period search performed using supercomputers at the Interdisciplinary
Centre for Mathematical and Computational Modeling of Warsaw University
(ICM UW). We searched for periodicity all 32 million stars in the LMC using
program {\sc Fnpeaks} by Z.~Ko{\l}aczkowski. We tested the range of
frequencies from 0.0 to 24.0 cycles per day, with a frequency step of
0.0001. For each star the ten highest peaks in the power spectrum were
recorded with appropriate amplitudes and S/N parameters. Then, the third
order Fourier series was fitted to each light curve folded with the
dominant period, the function was subtracted from the data, and the
procedure of period searching was repeated on the residual data.

The first criterion used for the identification of classical Cepheids in
our data were positions in the PL diagrams. We tested stars located not
only strictly in the PL relations for classical Cepheids, but also in wide
region above and below these sequences, including type II Cepheids which
will be published in a forthcoming paper. We used PL diagrams in various
wave bands: -- $I$, $V$, Wesenheit index, and near-infrared $K$ band from
the 2MASS project (Cutri \etal 2003).

Tens of thousands light curves selected in this manner were subsequently
subjected for a visual inspection. During the careful inspection the
variables were divided into pulsating-like stars, eclipsing binaries and
other variable objects. Then, the candidates for pulsating variables were
filtered according to the $(V-I)$ colors. We removed from the list objects
bluer than $(V-I)=0.2$~mag and redder than $(V-I)=1.8$~mag, however we
carefully checked, if the rejected objects are not real classical Cepheids
extraordinarily reddened or with erroneous photometry.

Stars which remained on the list were a mixture of various pulsating
variables crossing the classical instability strip: classical Cepheids,
type II Cepheids, anomalous Cepheids, RR~Lyr stars from the Galaxy and the
LMC and High Amplitude $\delta$~Sct stars (HADS). The long period classical
Cepheids were relatively easy to distinguish, due to their narrow PL
relations and characteristic shapes of the light curves. However, in the
short-period domain (for $P<3$~days and in particular for $P<1$~days) the
PL sequences overlap with various types of pulsating variables. Separation
of classical Cepheids from other pulsators was based mainly on the shapes
of their light curves. We will discuss this problem in the next part of the
OIII-CVS.

Our data show that first-overtone classical Cepheids and HADS follow the
same PL relation, with no discontinuity. Thus, our distinction between both
groups is arbitrary, being a matter of convention. To define limiting
period we used double-mode pulsators, which we detected in significant
number among classical Cepheids and HADS. In the Petersen diagram (\ie the
ratio of periods \vs logarithm of the longer period plot, Petersen 1973)
the stars pulsating simultaneously in the fundamental mode and first
overtone (F/1O) are naturally separated into two groups, with the gap
between (fundamental mode) periods in the range 0.4--1~days. Among
pulsators with the first two overtones excited (1O/2O) we detected only one
group with the shortest first overtone period of about 0.24 days. Thus,
$P_1=0.24$~days was taken for the first-overtone pulsators as a boundary
between HADS and classical Cepheids. For the fundamental-mode Cepheids we
cut our sample of $\delta$~Cep stars at $P_0=0.995$~days, \ie the shortest
F period of double-mode F/1O Cepheids (with exception of a peculiar object
OGLE-LMC-CEP-0083). We notice, that only a few pulsating-like variables
with shorter periods were found on the continuation of the fundamental-mode
Cepheids PL relation, but the shapes of their light curves were
considerably different than for F Cepheids.
The preliminary distinction between fundamental (F) and first overtone (1O)
classical Cepheids was done using their positions in the $W_I$--$\log{P}$
diagram, but the final classification utilized the Fourier parameters
$R_{21}$ and $\phi_{21}$ (see Section~6). The exemplary light curves of
single-mode Cepheids from the whole range of periods and luminosities
covered by OGLE are presented in Fig.~1.

\begin{figure}[htb]
\centerline{\includegraphics[width=13cm]{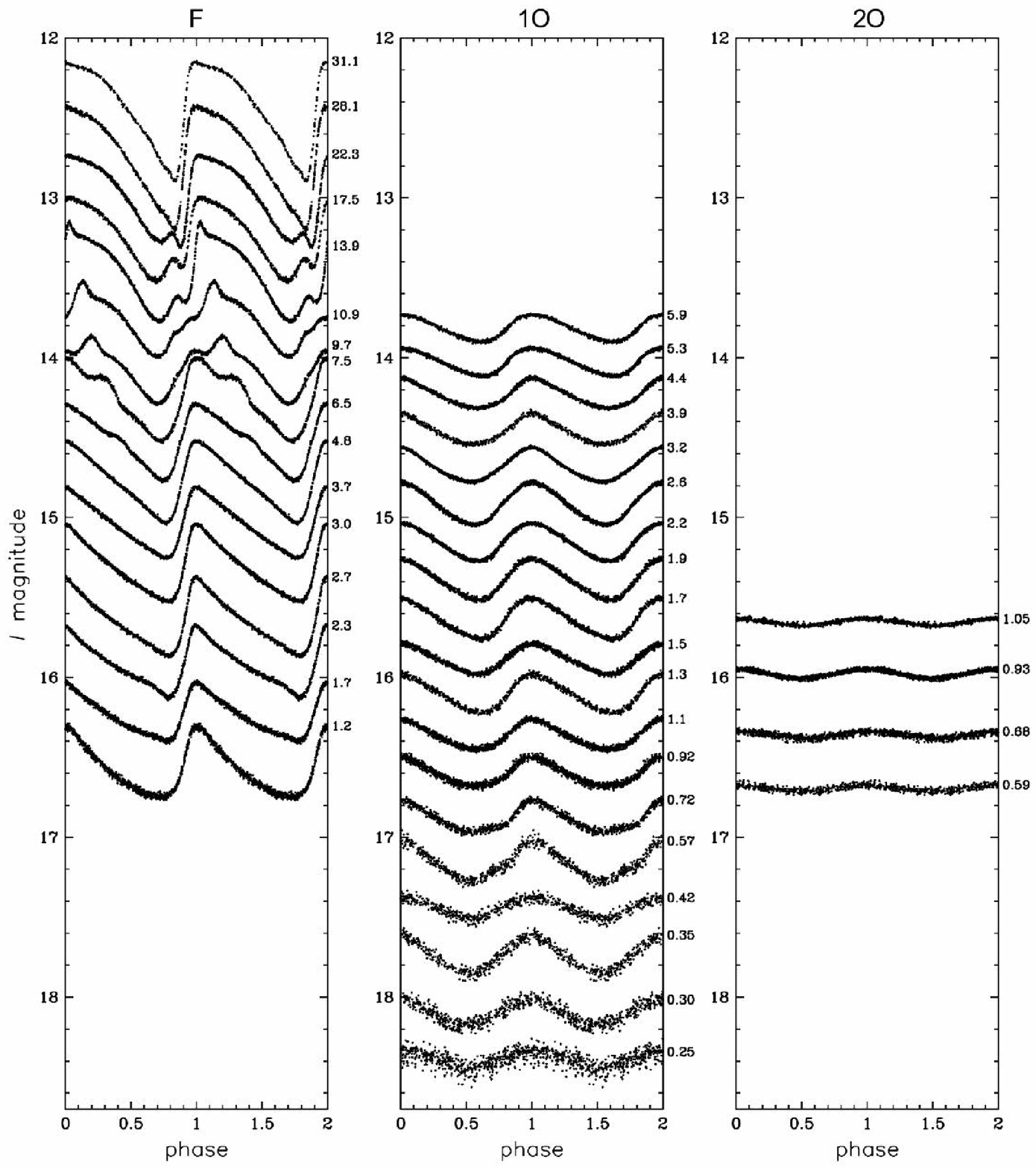}}
\FigCap{Illustrative light curves of fundamental-mode ({\it left panel}),
first-overtone ({\it middle panel}) and second-overtone ({\it right panel})
Cepheids. Small numbers at the right side of each panel show the rounded
periods in days of the light curves presented in panels.}
\end{figure}

\subsection{Second-Overtone Cepheids}
Classical Cepheids pulsating solely in the second overtone are very rare
but astrophysically interesting objects, because they can be used as an
independent test of pulsational and evolutionary models (Antonello and
Kanbur 1997, Bono \etal 2001). There is only one potential candidate for
the pure 2O Cepheid in the Galaxy -- V473~Lyr (HR~7308 -- Burki \etal
1986). Alcock \etal (1999a) undertook a search for single-mode 2O Cepheids
in the LMC using double-mode Cepheids pulsating in the first and the second
overtones as templates. They separated both modes in beat Cepheids and
studied the 2O variations. Singly-periodic second overtone Cepheids should
have nearly sinusoidal light curves, small amplitudes and mean luminosities
slightly higher than 1O Cepheids of the same periods. In the
color--magnitude diagram these stars should occupy the blue edge of the
instability strip. Alcock \etal (1999a) proposed one candidate for 2O
Cepheid in the LMC.

Udalski \etal (1999b) followed the same strategy as MACHO group for
Cepheids in the Small Magellanic Cloud (SMC). Theoretical investigations
suggest that metal-poor environments favor Cepheids pulsating purely in the
second overtone mode. As a result, Udalski \etal (1999b) found 13 firm
candidates for 2O pulsators -- the largest such sample detected to date.

We started a search for singly-periodic second overtone Cepheids using all
LMC stars in our database. We selected stars with S/N of periods larger
than 9 and located just above the period -- {\it I}-band magnitude relation
for 1O Cepheids (spreading from the upper edge of the PL sequence to
0.75~mag above this line). Then, the light curves were visually inspected
and obvious eclipsing binaries were rejected. At this stage we noticed a
distinct group of stars located at the blue edge of the instability strip
($0.2<(V-I)<0.6$~mag). From this group we removed a few pulsating variables
with asymmetrical light curves which we classified as blended RR~Lyr
stars. The last criterion of our selection was the ratio of amplitudes in
the $V$ and $I$ bands, what allowed us to remove a couple of ellipsoidal
variables from our list. Ellipsoidal modulation is mainly a geometrical
effect, so the amplitudes in two filters are very similar, while for
Cepheids, the {\it V}-band amplitudes are larger than for $I$ band by a
factor of about 1.7.

In total 14 objects passed our selection criteria, all of them in the
period range 0.58--1.2~days and in the low-amplitude domain. It is a very
homogeneous group which delineates additional PL sequence located above the
relation of the first-overtone pulsators. The shapes of the light curves
(see Fig.~1 for examples) are of the same type as in the 2O Cepheids found
by Udalski \etal (1999b), with Fourier parameter $R_{21}$ smaller
than~0.1. The MACHO candidate for 2O Cepheid in the LMC (Alcock \etal
1999a) is also on our list.

\subsection{Multiperiodic Cepheids}
Double-mode or beat Cepheids pulsate simultaneously in two radial modes --
usually fundamental mode and first overtone (F/1O) or first and second
overtones (1O/2O). Very few such objects had been known before the large
microlensing surveys era. The situation changed when the MACHO project
announced the discovery of 45 beat Cepheids in the LMC (Alcock \etal 1995),
increased later to 73 objects (Welch \etal 1997). Based on the data
collected during the OGLE-II project the OGLE group published samples of 93
double-mode Cepheids in the SMC (Udalski
\etal 1999a) and 76 such stars in the LMC (Soszy{\'n}ski \etal 2000).

The search for double-mode Cepheids in the LMC was carried out in two
manners. First, using periods determined for all stars in the LMC we
calculated the ratios of the principal periods found in two iterations of
the period searching and plotted the Petersen diagram. Then, we selected
all the stars located close to the expected position of the double-mode
pulsators in the LMC. These stars were visually inspected, and objects with
clear double-periodic signal were left on the list. This way not only beat
Cepheids were selected, but also double mode $\delta$~Sct and RR~Lyr stars,
which will be presented in the next parts of the OIII-CVS.

Second, additional search for double-mode variables was performed for all
previously selected Cepheids. Each light curve was fitted by a Fourier
series with a number of harmonics minimizing the $\chi^2$ per degree of
freedom and the function was subtracted from the observational data. The
residuals were searched for other periodic signals and, if detected, such a
candidate was marked for visual inspection. We discovered several new
double-mode Cepheids and a number of other multi-periodic variables, such
as Cepheids with periodicities very close to the dominant periods, Cepheids
in eclipsing binary systems, Cepheids blended with other variable stars.

\begin{figure}[htb]
\centerline{\includegraphics[width=11.5cm, bb=25 270 565 745]{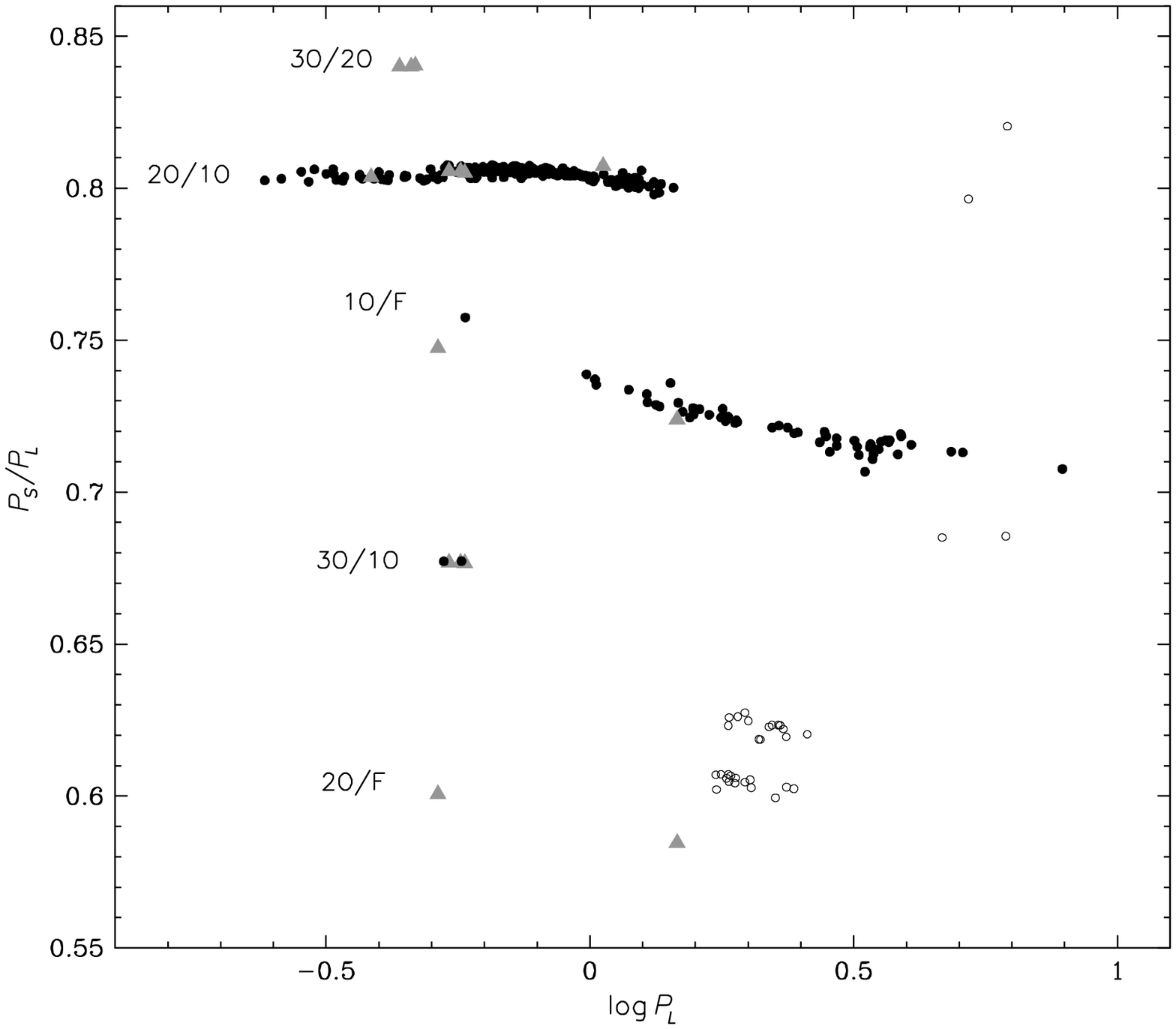}}
\FigCap{Petersen diagram for multiperiodic Cepheids in the LMC. Solid dots
represent double-mode (F/1O, 1O/2O and 1O/3O) Cepheids, grey triangles show
triple-mode Cepheids (three points per star) and empty circles show
selected other stars with significant secondary periods.}
\end{figure}
Fig.~2 shows the Petersen diagram for multi-mode Cepheids in our
list. Black points represent the Cepheids pulsating in two radial modes. It
is worth emphasizing that our sample of double-mode Cepheids covers much
larger range of periods than presented to date. We especially point out for
one long-period F/1O Cepheid OGLE-LMC-CEP-1082 with fundamental-mode period
equal to 7.86~days, however connected with exceptionally low amplitude of
pulsations. In the list of F/1O double-mode Cepheid we included also one
double-periodic pulsator, OGLE-LMC-CEP-0083, with exceptionally short
periods (0.581~d and~0.440~d) and the period ratio placing this object in
the Petersen diagram somewhat above the line connecting F/1O double-mode
Cepheids and HADS. On the other hand, we found a considerable number of
short-period 1O/2O Cepheids, with the first-overtone periods even as short
as 0.24~days. It is remarkable that in the range of 1O periods
0.5--0.75~days double-mode pulsators are significantly more common than
single-mode Cepheids.

We also performed a search for triple mode Cepheids. Three new stars of
that type were discovered in addition to two already known triple-mode
Cepheids in the LMC (Moskalik \etal 2004). The analysis of these stars is
presented in the paper by Soszy{\'n}ski \etal (2008). In the same work we
also announced the discovery of two double-mode Cepheids pulsating
simultaneously in the first and third overtone modes. This is a new class
of double-mode Cepheids.

During the search for multiperiodicity we detected a significant number of
Cepheids with the secondary periods very close to the primary ones. Such
ratios of periods close to 1 are usually connected with a long-term
amplitude and/or phase modulation, and, by analogy to RR~Lyr stars, these
objects are called Blazhko Cepheids. Using our simple analysis we detected
such behavior in about 4\% of F Cepheids, 28\% of 1O and in the same
fraction of 2O single-mode Cepheids, 18\% of F/1O beat Cepheids, and 36\%
of 1O/2O double-mode Cepheids. For single-mode pulsators this is somewhat
larger fraction than detected by Moskalik and Ko{\l}aczkowski (2008b) in
the OGLE-II data, what can be explained by the longer time span of our new
photometry. Among double-mode Cepheids, the fraction of non-radial
pulsators seem to be comparable with values determined by Moskalik and
Ko{\l}aczkowski (2008a,b).

We also noticed in the Petersen diagram a number of objects with period
ratios that do not match any known phenomena. Especially numerous group was
found for period ratios of 0.60--0.63~days and (longer) periods in the
range 1.7--2.6~days. It is interesting that the longer periods are always
related to the first overtone mode, while the shorter periods are connected
to low-amplitude variations. A few such Cepheids were recently listed by
Moskalik \etal (2008b) who suggested that such a behavior is connected with
a kind of non-radial oscillations. In our Petersen diagram these stars seem
to follow two sequences, with period ratios in the ranges 0.60--0.61 and
0.62--0.63.

\begin{figure}[htb]
\centerline{\includegraphics[width=14cm]{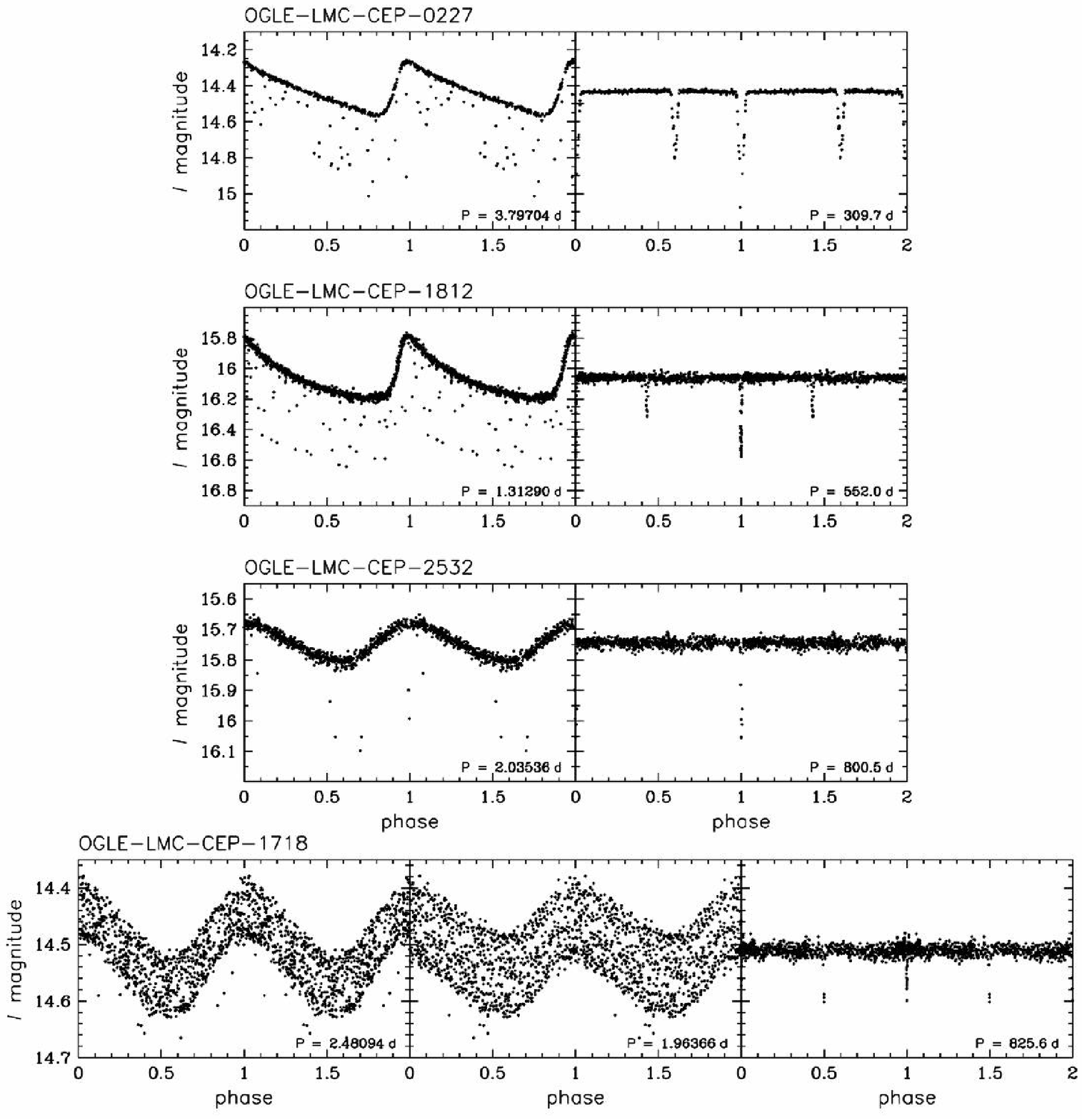}}
\FigCap{Light curves of Cepheids with additional eclipsing
variability. {\it Left panels} show the original photometric data folded
with the Cepheid periods. {\it Right panels} show the eclipsing light
curves after subtracting the Cepheid component. In the case of
OGLE-LMC-CEP-1718 two Cepheid light curves are presented.}
\end{figure}
As a by-product of searching for double-mode pulsators we detected three
Cepheids with eclipsing modulation imposed on the pulsational light curves:
OGLE-LMC-CEP-0227, OGLE-LMC-CEP-1812 and OGLE-LMC-CEP-2532. Only the last
one was already reported by Udalski \etal (1999d -- LMC\_SC16 119952). Light
curves of these stars are presented in Fig.~3. Cepheids with eclipsing
variations are of particular interest, because, if they are physical
binaries with the pulsating star as one of the components, they can be used
for direct determination of the mass and radius of Cepheid variable. In
some cases it is possible to distinguish between physically related objects
and optical binaries. If the eclipses are caused by a nearby star which is
not resolved by standard photometry, we should have found the shift in
centroid positions measured during eclipse and out of eclipse. No such
shift was found for these objects. Four other Cepheids --
OGLE-LMC-CEP-0388, OGLE-LMC-CEP-2052, OGLE-LMC-CEP-2095, and
OGLE-LMC-CEP-3037 -- are suspected to be in eclipsing binary systems. They
are brighter than other Cepheids with the same periods, but we detected
only a few points in each of these stars that can be associated with
potential eclipses.

We also draw attention to the extremely interesting object
OGLE-LMC-CEP-1718, which is a system consisting of two first overtone
Cepheids with eclipses visible every 412.8 days. This object was discovered
by Alcock \etal (1995 -- MACHO*05:21:54.8-69:21:50) and noticed by Udalski
\etal (1999d), but the eclipses were overlooked. Fig.~3 shows the light
curves of this object. Besides, we also detected two already known double
Cepheids (Alcock \etal 1995): OGLE-LMC-CEP-0571 and OGLE-LMC-CEP-0835.

There are also several Cepheids in our sample exhibiting additional
short-period and low-amplitude eclipsing modulation certainly caused by
blending. There are also Cepheids with additional variability of other
types, \eg cataclysmic-like changes, semiregular variations, long secondary
periods, etc. The majority of these cases can be explained as an effect of
physical or optical binarity with other variable star.

Since the time base of the OGLE-III observations is 2400 days, and for the
merged OGLE-II and OGLE-III data is longer than 4000 days, our data offer a
possibility of studying various long-term effects in the Cepheids -- period
changes, amplitude and phase modulation, or mean magnitude changes. The
full study of the period changing Cepheids will be published in the
forthcoming paper (Poleski 2008, in preparation).

\Section{Catalog of Classical Cepheids in the LMC}
In total 3361 classical Cepheids were found in the LMC OGLE-III fields. The
sample consists of 1848 fundamental-mode, 1228 first-overtone, 14
second-over\-tone, 61 double mode F/1O, 203 double mode 1O/2O, 2
double-mode 1O/3O, 2~triple-mode F/1O/2O and 3~triple-mode 1O/2O/3O
classical Cepheids.

In addition we prepared a list of 23 low amplitude variables which can be
related to classical Cepheids, because they lie in the vicinity of Cepheids
in the color--magnitude and PL diagrams. Such low-amplitude Cepheids are
expected by the evolutionary models as pulsators entering or exiting the
instability strip.
\begin{landscape}
\renewcommand{\TableFont}{\scriptsize}
\MakeTable{l@{\hspace{9pt}}
l@{\hspace{6pt}}
r@{\hspace{1pt}}
r@{\hspace{6pt}}
l@{\hspace{3pt}}
l@{\hspace{3pt}}
l@{\hspace{3pt}}
l@{\hspace{3pt}}
l@{\hspace{-1pt}}
c@{\hspace{-1pt}}
l}{12.5cm}{Exemplary part of the {\sf ident.dat} file}
{\hline
 \multicolumn{1}{c}{Cepheid ID} 
&\multicolumn{2}{c}{OGLE-III ID} 
&\multicolumn{1}{c}{Mode} 
&\multicolumn{1}{c}{RA} 
&\multicolumn{1}{c}{DEC} 
&\multicolumn{1}{c}{OGLE-II ID} 
&\multicolumn{1}{c}{MACHO ID} 
&\multicolumn{1}{c}{ASAS ID} 
&\multicolumn{1}{c}{GCVS ID} 
&Other \\
&\multicolumn{1}{c}{Field} 
&\multicolumn{1}{c}{No} & 
&\multicolumn{1}{c}{[J2000.0]} 
&\multicolumn{1}{c}{[J2000.0]} & & & & LMC... & designation \\
\hline
\multicolumn{11}{l}{...}\\
OGLE-LMC-CEP-0931 & LMC118.5 & 23110  &      F & 05:05:57.39 &$-$68:26:17.9 & LMC\_SC13\_125152 &            &                 & V1015 & HV2305 \\
OGLE-LMC-CEP-0932 & LMC116.6 & 12576  &     1O & 05:05:59.04 &$-$67:20:42.4 &                   &            &                 &       & \\
OGLE-LMC-CEP-0933 & LMC115.6 & 14990  &      F & 05:05:59.19 &$-$66:47:15.0 &                   &            &                 & V1001 & HV13058 \\
OGLE-LMC-CEP-0934 & LMC121.2 & 30142  &      F & 05:06:00.15 &$-$70:27:22.4 &                   &23.4151.25  &                 & V1041 & HV2330 \\
OGLE-LMC-CEP-0935 & LMC119.6 &101320  &      F & 05:06:00.89 &$-$69:06:17.1 & LMC\_SC13\_74156  &            &   050601-6906.3 & V1032 & HV893 \\
OGLE-LMC-CEP-0936 & LMC118.6 & 29152  &      F & 05:06:01.13 &$-$68:37:38.5 & LMC\_SC13\_111968 &            &                 &       & \\
OGLE-LMC-CEP-0937 & LMC117.7 & 49519  &   F/1O & 05:06:02.74 &$-$68:06:02.3 &                   &19.4186.876 &                 &       & \\
OGLE-LMC-CEP-0938 & LMC118.7 & 77581  &      F & 05:06:05.75 &$-$68:41:52.1 & LMC\_SC13\_106993 &            &                 & V1019 & HV13022 LMV1696 \\
OGLE-LMC-CEP-0939 & LMC115.7 & 14854  &      F & 05:06:05.98 &$-$66:59:43.9 &                   &            &                 & V1013 & HV13057 \\
OGLE-LMC-CEP-0940 & LMC121.2 & 36964  &      F & 05:06:07.06 &$-$70:26:08.8 &                   &23.4151.27  &                 & V1048 & DV81 \\
OGLE-LMC-CEP-0941 & LMC115.5 & 14135  &      F & 05:06:07.13 &$-$66:37:49.3 &                   &            &                 &       & \\
OGLE-LMC-CEP-0942 & LMC121.4 & 39539  &     1O & 05:06:08.12 &$-$70:11:16.7 &                   &            &                 &       & \\
OGLE-LMC-CEP-0943 & LMC122.1 &  3754  &      F & 05:06:08.88 &$-$71:15:26.1 &                   &            &   050609-7115.4 & V1059 & HV2338 \\
OGLE-LMC-CEP-0944 & LMC121.3 &  7423  &     1O & 05:06:11.39 &$-$70:25:45.7 &                   &23.4151.40  &                 &       & \\
OGLE-LMC-CEP-0945 & LMC115.5 & 17550  &      F & 05:06:14.81 &$-$66:40:45.9 &                   &            &   050615-6640.8 & V1022 & HV2294 \\
OGLE-LMC-CEP-0946 & LMC118.8 & 85656  &     1O & 05:06:14.89 &$-$68:51:56.7 & LMC\_SC13\_160626 &1.4175.14   &                 &       & \\
OGLE-LMC-CEP-0947 & LMC117.8 & 26519  &     1O & 05:06:14.94 &$-$68:19:03.1 & LMC\_SC13\_197484 &            &                 &       & \\
OGLE-LMC-CEP-0948 & LMC118.7 & 34808  &     1O & 05:06:16.43 &$-$68:47:20.5 & LMC\_SC13\_165223 &1.4176.34   &                 &       & \\
OGLE-LMC-CEP-0949 & LMC118.7 & 77545  &      F & 05:06:16.89 &$-$68:40:33.7 & LMC\_SC13\_173745 &19.4178.3   &                 & V1040 & HV895 \\
OGLE-LMC-CEP-0950 & LMC119.1 &    81  &     1O & 05:06:18.33 &$-$69:31:47.9 & LMC\_SC12\_133630 &            &                 &       & \\
OGLE-LMC-CEP-0951 & LMC115.5 & 14063  &      F & 05:06:18.53 &$-$66:38:44.7 &                   &            &                 & V1029 & W48 \\
OGLE-LMC-CEP-0952 & LMC115.5 & 14065  &      F & 05:06:19.60 &$-$66:35:47.8 &                   &            &                 &  & \\
OGLE-LMC-CEP-0953 & LMC122.4 &   139  &     1O & 05:06:22.14 &$-$70:48:05.3 &                   & 23.4146.20 &                 &  & \\
OGLE-LMC-CEP-0954 & LMC117.6 & 51980  &      F & 05:06:22.87 &$-$67:58:33.5 &                   & 19.4188.23 &                 & V1042 & HV5571 LMV779 \\
OGLE-LMC-CEP-0955 & LMC121.2 &  7798  &     1O & 05:06:29.11 &$-$70:31:03.4 &                   & 23.4150.47 &                 &  &  \\
\multicolumn{11}{l}{...}\\
\hline}
\end{landscape}
\noindent
Seven candidates for ultra-low amplitude Cepheids in the
LMC were found by Buchler \etal (2005). We stress that our sample of low
amplitude variables not necessarily consist of Cepheids only. Such light
curves can be a product of ellipsoidal or rotational modulation, or
blending with other type of variable star.

The OIII-CVS is available in the electronic version only from the OGLE
Internet archive:
\begin{center}
{\it http://ogle.astrouw.edu.pl/} \\ {\it
ftp://ftp.astrouw.edu.pl/ogle/ogle3/OIII-CVS/lmc/cep/}\\
\end{center}
The catalog is accessible through a user-friendly WWW interface or {\it
via} FTP site. In the FTP the full list of our sample of classical Cepheids
is given in the file {\sf ident.dat}. Part of this file is shown in
Table~1. All objects are arranged according to their right ascension. The
Cepheids have designations of the form OGLE-LMC-CEP-NNNN, where NNNN is a
four digit consecutive number. In the following columns of Table~1 we
provide: Cepheid ID, OGLE-III field and the database number of star
(consistent with the LMC photometric maps of Udalski \etal 2008b), mode of
pulsation, RA and DEC coordinates for the epoch 2000.0 and
cross-identifications of our objects with previously published catalogs of
Cepheids in the LMC (see Section~5), successively: OGLE-II, MACHO, ASAS and
General Catalogue of Variable Stars (GCVS). Last columns contain other
designations taken from the GCVS.

\renewcommand{\TableFont}{\scriptsize}
\MakeTableee{l@{\hspace{4pt}}
c@{\hspace{4pt}}
c@{\hspace{2pt}}
r@{\hspace{4pt}}
c@{\hspace{4pt}}
c@{\hspace{4pt}}
c@{\hspace{4pt}}
c@{\hspace{4pt}}
c@{\hspace{4pt}}
c@{\hspace{4pt}}c}{12.5cm}{First 20 lines of the {\sf cepF.dat} file}
{\hline
\noalign{\vskip3pt}
\multicolumn{1}{c}{Cepheid ID} & $\langle{I}\rangle$ & $\langle{V}\rangle$
&\multicolumn{1}{c}{$P$} & $\sigma_P$ & $T_{\rm max}$ & $A(I)$ & $R_{21}$ & $\phi_{21}$ & $R_{31}$ & $\phi_{31}$ \\
&[mag]& [mag] & \multicolumn{1}{c}{[days]} & [days] & {\tiny (HJD-2450000)}
&[mag]& & & & \\
\noalign{\vskip3pt}
\hline
\noalign{\vskip3pt}
OGLE-LMC-CEP-0002 & 15.672 &16.412 &  3.1181195 &0.0000161 & 2171.23921 & 0.257 & 0.296 &4.705 & 0.101 &2.962\\
OGLE-LMC-CEP-0005 & 14.661 &15.413 &  5.6120581 &0.0000135 & 2171.78078 & 0.521 & 0.431 &4.971 & 0.167 &3.391\\
OGLE-LMC-CEP-0012 & 15.469 &16.067 &  2.6601882 &0.0000022 & 2162.43751 & 0.688 & 0.522 &4.402 & 0.314 &2.611\\
OGLE-LMC-CEP-0016 & 13.707 &14.787 & 10.5064564 &0.0015553 & 2157.57180 & 0.115 & 0.028 &6.015 & 0.106 &0.923\\
OGLE-LMC-CEP-0017 & 15.345 &15.992 &  3.6772562 &0.0000299 & 2169.48050 & 0.611 & 0.494 &4.594 & 0.287 &2.983\\
OGLE-LMC-CEP-0018 & 15.222 &16.051 &  4.0478526 &0.0000275 & 2165.39166 & 0.265 & 0.284 &4.567 & 0.086 &2.998\\
OGLE-LMC-CEP-0021 & 14.722 &15.491 &  5.4579746 &0.0000259 & 2165.34485 & 0.431 & 0.416 &4.969 & 0.145 &3.555\\
OGLE-LMC-CEP-0023 & 16.325 &17.044 &  1.7018254 &0.0000019 & 2164.75709 & 0.403 & 0.441 &4.352 & 0.261 &2.493\\
OGLE-LMC-CEP-0025 & 15.343 &16.157 &  3.7334998 &0.0000099 & 2165.69444 & 0.360 & 0.409 &4.694 & 0.173 &3.131\\
OGLE-LMC-CEP-0026 & 15.466 &16.081 &  2.5706764 &0.0000033 & 2164.25802 & 0.622 & 0.442 &4.275 & 0.228 &2.263\\
OGLE-LMC-CEP-0027 & 15.039 &15.641 &  3.5229468 &0.0000052 & 2165.27241 & 0.647 & 0.449 &4.377 & 0.254 &2.509\\
OGLE-LMC-CEP-0028 & 16.620 &17.276 &  1.2629545 &0.0000007 & 2171.08146 & 0.513 & 0.475 &4.174 & 0.280 &2.120\\
OGLE-LMC-CEP-0033 & 14.400 &15.233 &  7.1807757 &0.0000217 & 2160.83200 & 0.488 & 0.339 &5.377 & 0.213 &3.723\\
OGLE-LMC-CEP-0034 & 13.737 &14.630 & 11.2546276 &0.0000595 & 2180.63850 & 0.478 & 0.170 &4.758 & 0.107 &5.295\\
OGLE-LMC-CEP-0035 & 14.375 &15.157 &  6.9436898 &0.0000448 & 2165.30738 & 0.393 & 0.359 &5.365 & 0.138 &3.768\\
OGLE-LMC-CEP-0037 & 15.525 &16.308 &  3.0669047 &0.0000082 & 2165.55271 & 0.271 & 0.345 &4.576 & 0.136 &2.958\\
OGLE-LMC-CEP-0039 & 15.415 &16.069 &  3.1477368 &0.0000291 & 2164.95088 & 0.338 & 0.411 &4.568 & 0.180 &2.783\\
OGLE-LMC-CEP-0040 & 14.662 &15.408 &  5.1651454 &0.0000080 & 2182.37433 & 0.574 & 0.436 &4.633 & 0.204 &3.009\\
OGLE-LMC-CEP-0041 & 15.522 &16.270 &  2.9106624 &0.0000235 & 2165.81479 & 0.181 & 0.245 &4.451 & 0.043 &2.401\\
OGLE-LMC-CEP-0042 & 15.673 &16.367 &  2.5770292 &0.0000034 & 2183.19453 & 0.599 & 0.485 &4.377 & 0.272 &2.568\\
\hline}
Files {\sf cepF.dat}, {\sf cep1O.dat}, {\sf cep2O.dat}, {\sf cepF1O.dat},
{\sf cep1O2O.dat},\linebreak {\sf cep1O3O.dat}, {\sf cepF1O2O.dat}, and
{\sf cep1O2O3O.dat} list basic parameters of the sin\-gle-, double- and
triple-mode Cepheids with the appropriate modes excited. For single-mode
objects the consecutive columns contain: intensity mean magnitudes in the
$I$ and $V$ bands, periods in days and their uncertainties, moments of the
zero phase corresponding to maximum light, amplitudes in the $I$-band, and
Fourier parameters $R_{21}$, $\phi_{21}$, $R_{31}$, $\phi_{31}$ derived for
the {\it I}-band light curves. For double-mode and triple-mode pulsators
the format of tables is longer including additional periodicities. First
rows of the file {\sf cepF.dat} are shown in Table~2.

The file {\sf remarks.txt} contains additional information on some
Cepheids. We provide here remarks about uncertain classification,
interesting features as additional variability, variable mean magnitudes or
amplitude modulation, and information about differences compared to other
catalogs (for example different periods). Directory {\sf phot/} contains
{\it I}- and {\it V}-band OGLE photometry of the stars in our catalog. If
available, OGLE-II data are merged with the OGLE-III data. Finally, the
directory {\sf fcharts/} contains finding charts of all objects. These are
the $60\arcs\times60\arcs$ segments of the {\it I}-band DIA reference
images, oriented with N up, and E to the left.

\begin{figure}[htb]
\centerline{\includegraphics[width=13cm]{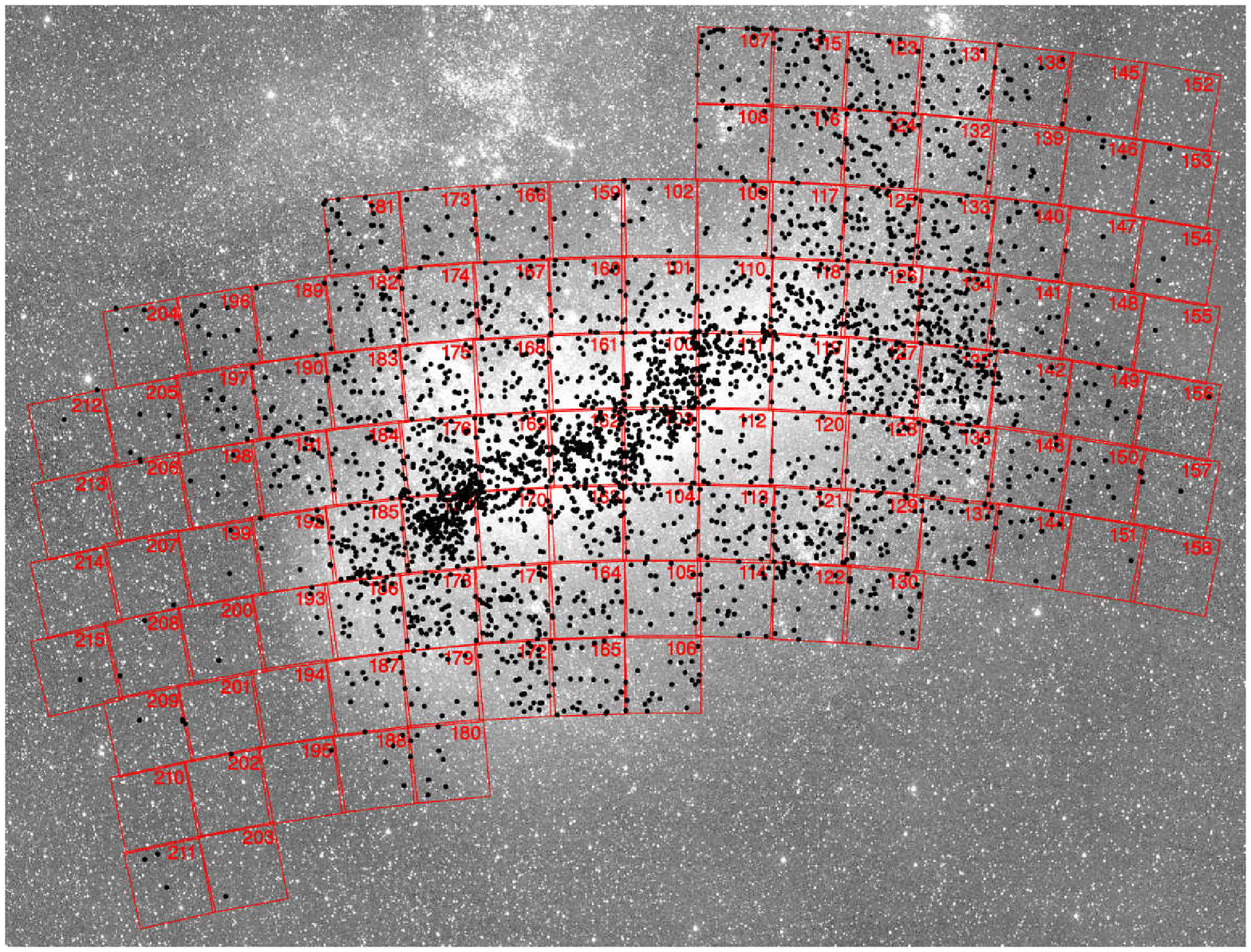}}
\vskip0.5cm
\FigCap{OGLE-III fields in the LMC. Dots indicate positions of classical
Cepheids from the OIII-CVS catalog. The background image of the LMC is
originated from the ASAS wide field sky survey.}
\end{figure}
Fig.~4 shows the spatial distribution of the classical Cepheids in the
LMC. The picture of the LMC was taken by the ASAS-3 wide field sky survey
(Pojma{\'n}ski 1997). Illustrative light curves of Cepheids pulsating in
the fundamental, first and second overtone modes are shown in Fig.~1. One
can notice the Hertzsprung progression (Hertzsprung 1926) for long period
fundamental-mode pulsators, \ie a~bump moving toward earlier phases with
increasing periods.

\Section{Cross-Identification with Previous Catalogs}
The catalog includes matches of our objects with previously published list
of classical Cepheids in the LMC. We queried the following catalogs:
OGLE-II catalogs of Cepheids in the LMC (Udalski \etal 1999d, Soszy{\'n}ski
\etal 2000), on-line MACHO Variable Star
Catalog\footnote{{\it http://wwwmacho.mcmaster.ca/}}, the extragalactic
part of the General Catalogue of Variable Stars (Artyukhina \etal 1995),
and ASAS Catalog of Variable Stars\footnote{{\it
http://www.astrouw.edu.pl/asas/}} (Pojma{\'n}ski 2002). In our sample of
Cepheids 2367 stars were already published in any of these catalogs, 994
objects are new detections.

With the aim of testing the completeness of our catalog we carefully
checked all objects not present in our list, but potentially covered by the
OGLE-III fields. Compared to the OGLE-II catalog of Cepheids in the LMC
published by Udalski \etal (1999d -- single-mode Cepheids) and
Soszy{\'n}ski \etal (2000 -- double-mode Cepheids) we missed 8 classical
Cepheids (not counting 16 objects saturated in the OGLE-III data). Five of
these Cepheids are members of LMC clusters and severe crowding affected the
OGLE-III photometry. The three remaining objects were close to the edges of
the fields and were affected by a small number of observations. We
supplemented our catalog with these missing objects.

The MACHO project released the list of about 1800 Cepheids in the
LMC. 1721~of them could potentially be found in the OGLE-III fields. We did
not find counterparts for 15 variables. Lack of five of these stars can be
explained by a small number of points at the field edges or problems with
the photometry in dense regions of the sky. The other objects were
classified by us as different type of variable stars, usually eclipsing
binaries.

The General Catalogue of Variable Stars contains 873 stars classified as
DCEP or DCEPS and potentially present in our fields in the LMC. We
performed an extensive searching for the counterparts of these stars in our
sample. In a number of cases the cross-identification between our sample
and the GCVS was uncertain, because of a large discrepancy (sometimes
larger than 2\arcm) in coordinates of the objects. Sometimes, the stars
were positionally coincident, but the periods disagreed. For many of these
objects we noticed that the period provided in the GCVS was an alias of the
true period. One missing object, LMC V0477 (DV53), we classified as
eclipsing or ellipsoidal binary. Another star, LMC~V0857 (HV~2284), was
recognized as Galactic RR~Lyr star. LMC~V0566 (HV~12509) is likely
a~type~II Cepheid in the eclipsing binary system.

We found no counterparts for nine stars classified as classical Cepheids in
the GCVS, namely LMC~V0224 (HV~12496), V0452 (DV44), V1620 (HV~13018),
V2175 (HV~5767), V2329, V2368 (HV~13032), V3972, V4366, and V4441
(HV\linebreak 12909). Since periods and coordinates provided by the GCVS
are sometimes erroneous, it is possible that these objects are present in
our sample, but each case should be directly checked using finding charts
from the literature. Such investigation should also reveal possible
Cepheids which stopped pulsating.

\Section{Basic Parameters}
The periods of variable stars and uncertainties of periods provided in the
catalog were calculated with the {\sc Tatry} program using multiharmonic
periodogram of Schwarzenberg-Czerny (1996). To determine mean luminosities,
amplitudes and Fourier parameters, each of the light curve was fitted by a
Fourier series of the order depending on the shape and scatter of the light
curve. The number of harmonics (maximum 12) was adjusted to minimize the
value of $\chi^2$ per degree of freedom. In this procedure we used the
program {\sc J-23} written by T.~Mizerski.

\begin{figure}[p]
\centerline{\includegraphics[width=13.5cm]{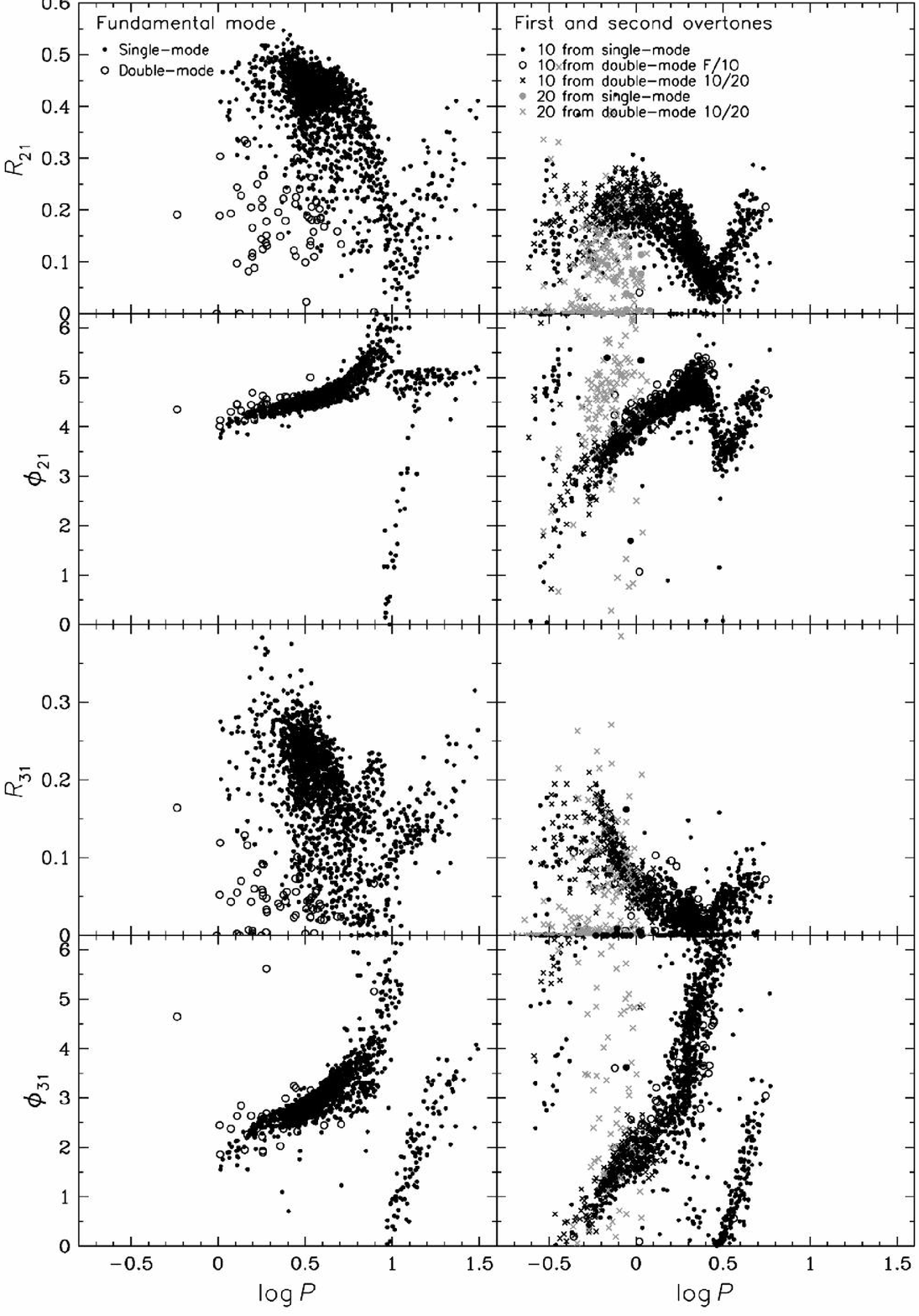}}
\FigCap{Fourier parameters of the fundamental-mode ({\it left panels}) and
overtone ({\it right panels}) Cepheids.}
\end{figure}

The {\it VI} intensity mean magnitudes were derived by integrating the
light curves converted to intensity units and transforming the result back
to the magnitude scale. The amplitudes provided with the catalog are the
differences between the maximum and minimum values of the function fitted
to the light curves. Fourier parameters -- amplitude ratios
$R_{k1}=A_k/A_1$ and phase differences $\phi_{k1}=\phi_k-k\phi_1$ (Simon
and Lee 1981) -- were derived using the same method of Fourier
decomposition. For light curves with insignificant higher harmonics the
amplitude ratios are equal to zero, while the appropriate phase differences
are not defined.

Fig.~5 shows Fourier parameters $R_{21}$, $\phi_{21}$, $R_{31}$, and
$\phi_{31}$ of the {\it I}-band light curves plotted against $\log{P}$. For
clarity fundamental-mode and overtone Cepheids are presented in separate
panels. Fourier coefficients are widely used tool for quantitative
description of the structure of Cepheid light curves. Complex pattern
visible in the diagrams reflects the Hertzsprung progression. The minimum
of $R_{21}$ at $P\approx10$~days for fundamental-mode Cepheids is
interpreted as a signature of 2:1 resonance between the fundamental and the
second overtone mode of pulsation (Andreasen and Petersen 1987). In the
vicinity of 10~days periods the $\phi_{21}$ parameter rises sharply to
$2\pi$ and appears again in the lower part of the diagram, what is caused
by rotation of $\phi_{21}$ modulo $2\pi$. Note that such sharp feature in
$\phi_{21}$ is not observed for some of the Cepheids around this period,
but all of them cross the line $\phi_{31}=0$ at $P\approx10$~days.

Similar behavior seem to appear twice for the first-overtone pulsators --
at periods $\approx0.35$~days and $\approx3$~days. The second feature is
interpreted as the signature of 2:1 resonance between the first and fourth
overtones (Antonello and Poretti 1986). The short-period discontinuity can
be explained by a presence of the 2:1 resonance between the first and fifth
overtones in stars with masses of about 2.5~\MS\ (W.~Dziembowski, private
communication).

\Section{Period-Luminosity Relation}
Period--luminosity relation of classical Cepheids plays a crucial role as
an indicator of the cosmic distance scale. The LMC is of special interest
in this field, because the extragalactic distances are calibrated with the
distance to this galaxy. Classical Cepheids found during the second phase
of the OGLE survey in the LMC were widely used in various programs aiming
at distance determinations, \eg HST Key Project (Freedman \etal 2001) or
Araucaria Project (Gieren \etal 2004).

The PL diagrams in the $V$ and $I$ magnitudes and in the reddening-free
Wesenheit index $W_I=I-1.55(V-I)$ (Madore 1982) are shown in Fig.~6. It is
striking that substantial fraction of points in the first two plots are
located considerably below the mean PL relations, but these stars generally
follow the narrow sequences in the period -- Wesenheit index plane. We
interpret such behavior as an effect of interstellar extinction -- very
variable from star to star. Considerable reddening for some Cepheids is
visible on the color--magnitude diagram plotted in Fig.~7.
\begin{figure}[p]
\centerline{\includegraphics[width=13.5cm, bb=35 55 545 745]{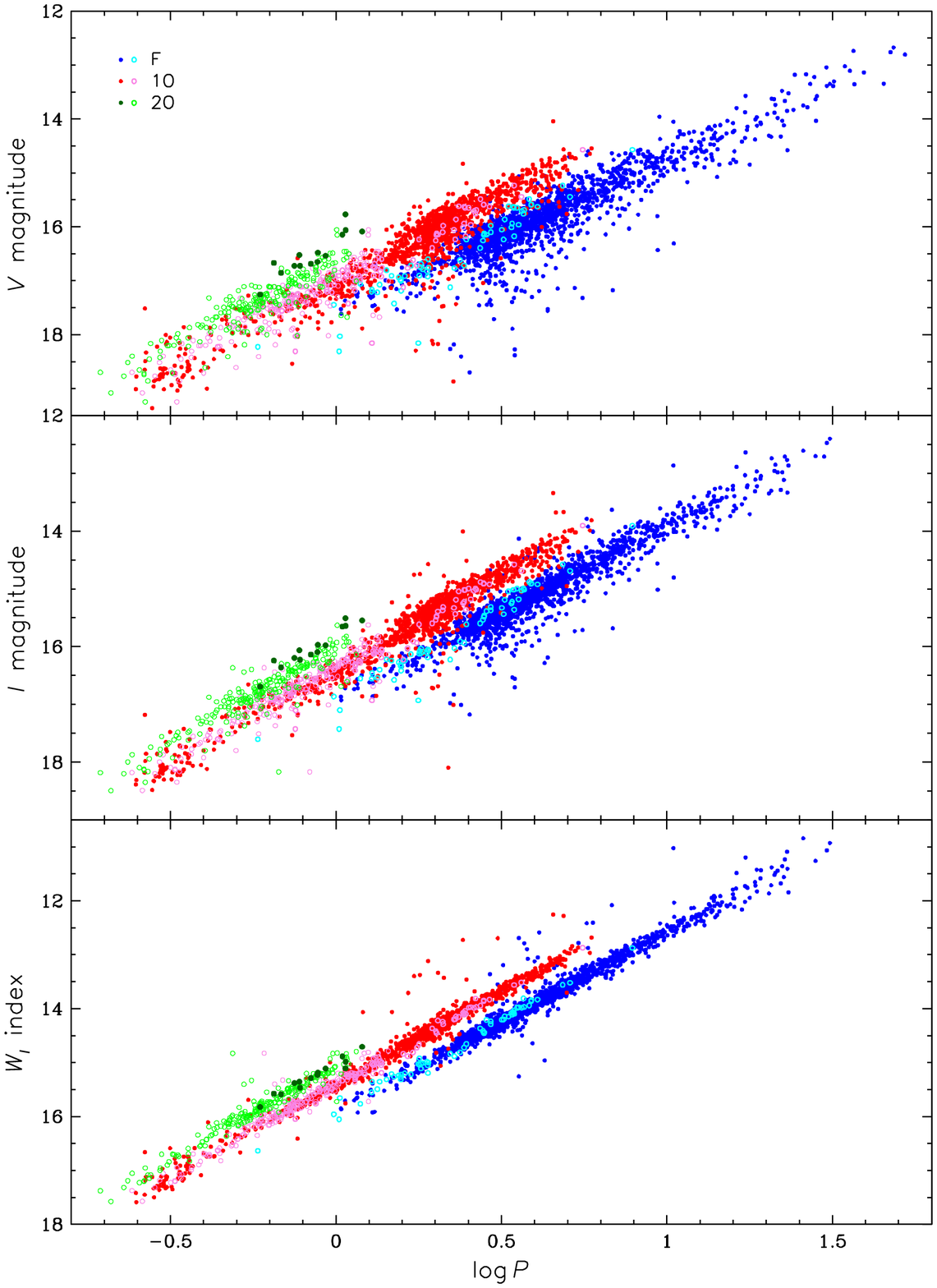}}
\FigCap{Period--luminosity diagrams for classical Cepheids in the LMC. Blue
and cyan points show fundamental-mode pulsators, red and magenta --
first-overtone, green -- second overtone. Solid dots are single-mode
Cepheids, while empty circles represent double-mode pulsators (two points
per star).}
\end{figure}
\begin{figure}[htb]
\centerline{\includegraphics[width=12.7cm]{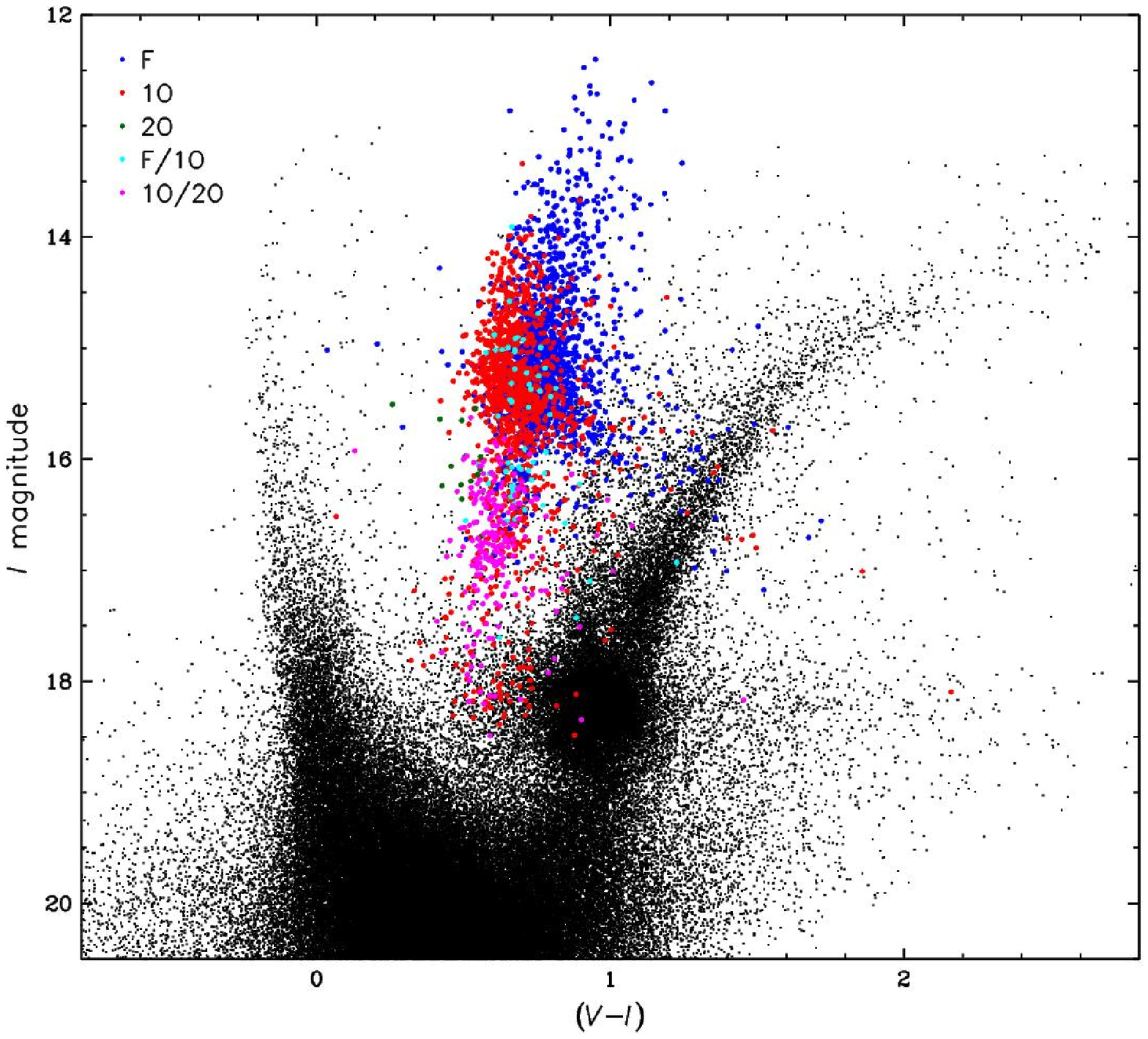}}
\FigCap{Color--magnitude diagram for classical Cepheids in the LMC. In the
background stars from the subfield LMC100.1 are shown. The significant
number of very red Cepheids are clearly located to the red side of the
respective instability strips for the various pulsation modes indicating
that large reddening is not unusual in the LMC.}
\end{figure}

We also carefully checked the stars that do not obey the PL relations in
the $\log{P}$--$W_I$ plane. In the majority of these cases this discrepancy
can be explained by blending with other star unresolved in our data. Most
of the Cepheids with superimposed additional type of variations (including
Cepheids with eclipsing modulations) do not match the period-$W_I$
laws. Notice that the PL sequence for the longest-period fundamental-mode
Cepheids seem to be more scattered than for the shorter-period
variables. It may be due to saturation effects, but the definitive answer
will be given after publication of the OGLE shallow survey to the LMC.

The PL relations provided below are not compensated for interstellar
extinction. We leave it to the reader to de-redden magnitudes using either
average extinction correction for the whole LMC or individual values for
each object. We also take no account for possible break of the linearity
suggested for the PL relation of fundamental mode Cepheids (see Ngeow \etal
2008 and references therein). We expect that our catalog will be used many
times to establish PL relations, and various subtle effects will be taken
into consideration.

The least squares solution with 3$\sigma$ clipping yields the following PL
relations for single-mode fundamental-mode classical Cepheids:
\begin{eqnarray*}
V  &=&-2.762(\pm0.022)\log{P}+17.530(\pm0.015)\qquad\sigma=0.22~{\rm mag} \\
I  &=&-2.959(\pm0.016)\log{P}+16.879(\pm0.010)\qquad\sigma=0.15~{\rm mag} \\
W_I&=&-3.314(\pm0.009)\log{P}+15.893(\pm0.006)\qquad\sigma=0.08~{\rm mag} \\
\end{eqnarray*}

\vspace*{-0.5cm}
\noindent and for the first-overtone mode:
\begin{eqnarray*}
V  &=&-3.194(\pm0.026)\log{P}+17.046(\pm0.009)\qquad\sigma=0.23~{\rm mag} \\
I  &=&-3.297(\pm0.017)\log{P}+16.405(\pm0.006)\qquad\sigma=0.16~{\rm mag} \\
W_I&=&-3.451(\pm0.009)\log{P}+15.398(\pm0.003)\qquad\sigma=0.07~{\rm mag.} \\
\end{eqnarray*}

\vspace*{-0.5cm}
The slopes of the PL relations for fundamental-mode Cepheids agree within
1$\sigma$ with previous determinations by Udalski \etal
(1999c)\footnote{updated coefficients are available from:\\ {\it
ftp://ftp.astrouw.edu.pl/ogle/ogle2/var\_stars/lmc/cep/catalog/README.PL}},
and Fouqu{\'e} \etal (2007). Of course, in the $V$ and $I$ domains we can
only compare the slopes, because we did not applied the reddening
correction, but in the $W_I$ extinction-free index we can consider both,
the slope and the zero point, of the PL relation. There is larger
discrepancy ($>2\sigma$) in the zero point of the $\log{P}$--$W_I$
relation. We also obtained somewhat larger scatter of the PL relations in
the OGLE-III sample, what can be explained by the inclination of the LMC
disk in respect to the line of sight. The OGLE-II fields covered only the
central parts of the LMC, what kept this effect much smaller. It is
important to note that Udalski \etal (1999c) and Fouqu{\'e} \etal (2007)
used the same definition of the Wesenheit index as we did, but this
definition depends on the adopted reddening law. Using different
coefficient in the Wesenheit index results in different slope and zero
point of the fitted PL relation.

The agreement in slopes is much worse for first-overtone Cepheids, although
still within $3\sigma$. We suspect that a considerable number of very
short-period first-overtone pulsators, not present in the OGLE-II sample of
Cepheids, which changed the fitted function could be an origin of this
discrepancy. The non-linearity of the PL relation may take place for the
first-overtone Cepheids, with the break at $P_1\approx0.5$~days.

\Section{Summary}
In this paper we present the largest catalog of classical Cepheids in the
LMC and probably the largest sample of such stars identified to date in any
environment. Our list of Cepheids is supplemented by the high quality,
long-term standard photometry enabling precise analysis of these
stars. These data are ideal for studying many fundamental problems, such as
interpretation of the pulsational and evolutionary models of Cepheids,
non-radial oscillations in the pulsating stars, possible non-linearity of
the PL relation, structure and history of the LMC.

The catalog contains very rare objects, such as Cepheids with three radial
modes excited, 1O/3O double-mode Cepheids, single-mode second-overtone
pulsators, Blazhko Cepheids, eclipsing binary systems containing Cepheids
including system of two Cepheids eclipsing each other. Our data show that
first-overtone classical Cepheids and high amplitude $\delta$~Sct stars
follow continuous PL relation. Distribution of the Fourier parameters
suggests that the internal resonance between radial modes may occur twice
for the first-overtone pulsators: for periods of about 0.35~days and
3~days. The PL relation for first-overtone Cepheids is possibly non-linear,
with a~discontinuity in the slope around $P=0.5$~days.

In the next parts of the OIII-CVS we will present other members of the
Cepheid family: type II Cepheids, anomalous Cepheids, HADS and RR~Lyr
stars. It is possible that the list of classical Cepheids described in this
paper will be supplemented with additional objects of this type detected
during further analysis.

\Acknow{The authors wish to thank Prof.~W.A.~Dziembowski, Prof.~M.~Feast,
and Prof.~W.~Gieren for many helpful suggestions which improved the paper.
We thank Drs. Z.~Ko{\l}aczkowski, T.~Mizerski, G.~Pojma{\'n}ski,
A.~Schwar\-zenberg-Czerny and J.~Skowron for providing the software and
data which enabled us to prepare this study.

This work has been supported by the Foundation for Polish Science through
the Homing (Powroty) Program and by MNiSW grants: NN203293533 to IS and
N20303032/4275 to AU.

The massive period searching was performed at the Interdisciplinary Centre
for Mathematical and Computational Modeling of Warsaw University (ICM
UW). We are grateful to Dr. M.~Cytowski for helping us in this analysis.}

\end{document}